\documentstyle[11pt]{article}
\newfont{\larom}{cmbx9 scaled\magstep3}
\newfont{\bsan}{cmssbx10}
\begin{document}
%%%%%%%%%%%%%%%%%%%%%%%%%%%%%%%%%%%%%%%%%%%%%%%%% Title and Abstract
\begin{center}
  {\larom The Apparent Fractal Conjecture\footnote{Communication presented
  as part of the talk delivered at the {\it ``South Africa Relativistic
  Cosmology Conference in Honour of George F.\ R.\ Ellis 60th Birthday''},
  University of Cape Town, February 1-5, 1999.}}\\
  \vspace{10mm}
  {\Large Marcelo B.\ Ribeiro}\\
  \vspace{6mm}
  Dept.\ Mathematical Physics, Institute of Physics,\\Federal University of
  Rio de Janeiro - UFRJ,\\C.P.\ 68528, Ilha do Fund\~ao, Rio de Janeiro, RJ
  21945-970,\\Brazil; E-mail: mbr@if.ufrj.br\\
  \vspace{10mm}
  {\bf ABSTRACT}
\end{center}
\begin{quotation}
  \small \it
  \noindent This short communication advances the hypothesis that the
  observed fractal structure of large-scale distribution of galaxies is
  due to a geometrical effect, which arises when observational
  quantities relevant for the characterization of a cosmological fractal
  structure are calculated along the past light cone. If this hypothesis
  proves, even partially, correct, most, if not all, objections raised
  against fractals in cosmology may be solved. For instance, under this
  view the standard cosmology has zero average density, as predicted by
  an infinite fractal structure, with, at the same time, the cosmological
  principle remaining valid. The theoretical results which suggest this
  conjecture are reviewed, as well as possible ways of checking its
  validity.
\end{quotation}
%%%%%%%%%%%%%%%%%%%%%%%%%%%%%%%%%%%%%%%%%%%%%%%%% pacs
\begin{flushleft} PACS numbers: 98.80.-k % cosmology
                       \ \ \    98.65.Dx % superclusters; large-scale structure of the Universe
                       \ \ \    98.80.Es % observational cosmology
                       \ \ \    05.45.Df % fractals 
\end{flushleft}
\vspace{10mm}
%%%%%%%%%%%%%%%%%%%%%%%%%%%%%%%%%%%%%%%%%%%%%%%%% Body
The issue of whether or not the large-scale distribution of matter in
the Universe actually follows a fractal pattern has divided cosmologists
in the last decade or so, with the debates around this thorny question
leading to a split of opinions between two main, and opposing, groups.

On one side, the {\it orthodox view} sustains that since a fractal
structure is inhomogeneous, it cannot agree with what we know about the
structure and evolution of the Universe, as this knowledge is based on the
cosmological principle and the Friedmann-Lema\^{\i}tre-Robertson-Walker
(FLRW) spacetime, with both predicting homogeneity for the universal
distribution of matter. Moreover, inasmuch as the cosmic microwave
background radiation (CMBR) is isotropic, a result predicted by the
FLRW cosmology, this group is, understandably, not prepared to give
up the standard FLRW universe model and the cosmological principle, as
that would mean giving up most, if not all, of what we learned about
the structure and evolution of the Universe since the dawn of cosmology
(Peebles 1993; Davis 1997; Wu, Lahav and Rees 1999; Mart\'{\i}nez 1999).

On the other side, the {\it heterodox view} claims that the systematic
and, under their methodology, unbiased interpretation of astronomical
data continuously shows that the distribution of galaxies is not
homogeneous, being in fact completely inhomogeneous up to present
observational scales, without any sign of homogenization. Therefore,
this other group not only disputes the traditional interpretation of 
astronomical data, but also the validity of the cosmological principle,
going as far as to implicitly suggest that the CMBR may be a cosmological
puzzle (Pietronero 1987; Coleman and Pietronero 1992; Pietronero,
Montuori and Sylos-Labini 1997; Sylos-Labini, Montuori and Pietronero
1998).

To reach those opposing conclusions, the validity of the methods used 
by both sides of this {\it ``fractal debate''} are, naturally, hotly
disputed, and so far there has not yet been achieved a consensus on this
issue. However, even if one is prone to part of the orthodox argument,
{\it i.e.}, that we cannot simply throw away some basic tenets of
modern cosmology, like the cosmological principle and the highly
successful FLRW cosmological model, when one looks in a dispassionate
way at the impressive data presented by the heterodox group, one cannot
dispel a certain uneasy feeling that something might be wrong in the
standard observational cosmology: the results are consistent and agree
with one another (Coles 1998).

Another interesting aspect to note about this fractal debate is that it
is as old as modern cosmology itself. The first suggestion that the
Universe could be constructed in a hierarchical, or fractal, manner dates
back to the very beginning of cosmology (Fournier D'Albe 1907; Charlier
1908, 1922), with contributions made even by Einstein (see Mandelbrot
1983, Ribeiro 1994, and references therein). Consequently, what we are
witnessing now is only the latest chapter of this old debate, which is
now focused on the statistical methods used by cosmologists to study
galaxy clustering. The previous chapter was between de Vaucouleurs
(1970ab) and Wertz (1970, 1971; see also de Vaucouleurs and Wertz 1971)
on one side, and Sandage, Tammann, and Hardy (1972) on the other side,
and was mainly focused on measurements of galaxy velocity fields and
deviations from uniform expansion, a topic which has also resurfaced
in the recent debate (Coles 1998). Therefore, it is clear that despite
being dismissed many times as ``unrealistic'', the fractal, or
hierarchical, concept has so far refused to die, being able to pass
from one generation of cosmologists to another (Oldershaw 1998). Thus,
considering its old history, and its incredible ability to survive, it
is perhaps premature to say that we are about to see this issue being
settled with the dismissal of the fractal concept.

From the brief summary above, it is clear that the two sides of the
fractal debate are locked in antagonistic and self-excluding
viewpoints. Nevertheless, it is the opinion of this author that this
divide may be not as radical as presented by both sides, and that it
is possible to build a bridge between both opinions, reconciling them 
by means of a change in perspective regarding how we deal with
observations in cosmology. What I intend to show next is that there is
already enough theoretical evidence to suggest that fractality can
be accommodated {\it within} the standard cosmology, where it would
appear as a real observational effect of {\it geometrical} nature,
arising from the way we carry out observations of large-scale structure
of the Universe. At the same time the cosmological principle, uniform
Hubble expansion, CMBR isotropy, and well defined meanings for the
cosmological parameters, such as $\Omega$, can survive, together with
the {\it observational fractality} obtained by the heterodox group
mentioned above. This perspective has the advantage of preserving most
of what we learned with the standard FLRW cosmology, and, at the same
time, making sense of Pietronero and collaborators' data, which, as
seen above, can no longer be easily dismissed.

The key point to understand how that can come about, is by
re-discussing the meaning of observations in cosmology. In relativistic
cosmology astronomical observations occur {\it along the past light
cone}, a fact which is often overlooked when one carries out
astronomical data reduction in cosmological models. Observers often
use cosmological formulae which does not take this key theoretical
feature into consideration. They are often under the assumption that at
the scales where observations are being made ($z < 1$) one can safely 
ignore that, since this is the region where the Hubble law is very
linear. However, Hubble law linearity has a range which does not
coincide with a constant density (see below), and the observed average
density is a key physical quantity for fractal characterization
(Pietronero 1987; Pietronero, Montuori and Sylos-Labini 1997; Coleman
and Pietronero 1992; Sylos-Labini, Montuori and Pietronero 1998;
Ribeiro and Miguelote 1998).

The first theoretical evidence that average density departs from local
homogeneity at much lower values for the redshift $z$ appeared in
Ribeiro (1992b). There, observational relations were calculated along
the past light cone
for unperturbed Einstein-de Sitter cosmology, and it was clearly
showed that deviations from local homogeneity start to occur at
$z \approx 0.04$, becoming very strong at $z \approx 0.1$. A plot of the
average density $\langle \rho \rangle$ against the luminosity distance
$d_\ell$ showed a continuous decrease in the average density, although
not in a linear manner. Still, Ribeiro (1992b) showed too that in the
Einstein-de Sitter cosmology the following limit holds,
\[ \lim_{d_\ell \rightarrow \infty} \langle \rho \rangle =0. \]

Many years ago, Wertz stated that a pure hierarchical cosmology ought
to obey the {\it ``zero global density postulate}: for a pure hierarchy
the global density exists and is zero everywhere'' (Wertz 1970, p.\ 18).
Such a result was also speculated by Pietronero (1987) as a natural
development of his fractal model. Therefore, what the above limit 
tells us is that the Einstein-de Sitter model {\it does} obey this key
requirement of fractal cosmologies. In addition, the decay of the
average density at increasing distances, another key aspect of a fractal
model, is also obeyed by the Einstein-de Sitter cosmology. Notice that
these two fractal features, present in all standard cosmological
models (Ribeiro 1993, 1994), appear {\it without any violation} of the
cosmological principle, linearity of Hubble law, and CMBR isotropy.
Moreover, cosmological parameters such as $q_0$, $\Omega_0$, $H_0$
still have their usual definitions and interpretations.

What is clear from Ribeiro (1992b), and sequel papers (Ribeiro 1993,
1994), was that the homogeneity of the standard cosmological models is
{\it spatial}, that is, it is a {\it geometrical} feature which does
not necessarily translate itself into an astronomically observable
quantity. Although a number of authors are aware of this fact, what
came as a surprise had been the calculated low redshift value where
this observational inhomogeneity appears. Therefore, it was clear by
then that relativistic effects start to play an important role in
observational cosmology at much lower redshift values than previously
assumed.

In another paper (Ribeiro 1995), the results above were further analysed
and it became clear why there seems to be no contradiction between
strong {\it observational} inhomogeneity and the linearity of the Hubble
law for $0.1 \le z < 1$ in Einstein-de Sitter cosmology. Due to the
non-linearity of the Einstein field equations, observational relations
behave differently at different redshift depths. Consequently, while the
linearity of the Hubble law is well preserved in the Einstein-de Sitter
model up to $z \approx 1$, a value implicitly assumed as the lower
limit up to where relativistic effects could be safely ignored,
the observational density is strongly affected by relativistic effects
at much lower redshift values. A power series expansion of these
two quantities showed that while the zeroth order term vanishes in the
distance-redshift relation, it is non-zero for the average density
as plotted against redshift. This zeroth order term is the main
factor for the different behaviour of these two observational
quantities at small redshifts. Pietronero, Montuori and Sylos-Labini
(1997) called this effect as the ``Hubble-de Vaucouleurs paradox''
(see also Baryshev et al.\ 1998). However, from the analysis
presented by Ribeiro (1995) it is clear that this is not a paradox,
but just very different relativistic effects on the observables at
the moderate redshift range ($0.1 \le z < 1$).

This effect explains why Sandage, Tammann and Hardy (1972) failed to
find deviations from uniform expansion in a hierarchical model: they
were expecting that such a strong inhomogeneity would affect the velocity
field, but it is clear now that if we take a relativistic perspective
for these effects they are not necessarily correlated at the range
expected by Sandage and collaborators. Notice that de Vaucouleurs and 
Wertz also expected that their inhomogeneous hierarchical models would
necessarily affect the velocity field, and change the linearity of the
Hubble law at $z <1$, and once such a change was not observed by Sandage,
Tammann, and Hardy (1972) it was thought that this implied an immediate
dismissal of the hierarchical concept. Again, this is not necessarily
the case if we take a relativistic view of those observational
quantities.

The results discussed above show that some key fractals features can
already be found in the simplest possible standard cosmological model,
that is, in the unperturbed Einstein-de Sitter universe. However, as
the average density decay is not linear in this model, considering all
these aspects we may naturally ask whether or not a perturbed model
could turn the density decay at increasing redshift depths into a
power law type decay, as predicted by the fractal description of
galaxy clustering. If this happens, then standard cosmology can be
reconciled with a fractal galaxy distribution. Notice that there are
some indications that this is a real possibility, as Amendola (1999)
pointed out that locally the cold dark matter and fractal models predict
the same behaviour for the power spectrum, a conclusion apparently shared
by Cappi et al.\ (1998). In addition, confirming Ribeiro's (1992b, 1995)
conclusions, departures from the expected Euclidean results at small
redshifts were also reported by Longair (1995, p.\ 398), and the starting
point for his findings was the same as employed by Ribeiro (1992b,
1995): the use of source number count expression along the null cone.

Considering all results outlined above, I feel there is enough grounds
to advance the following conjecture: {\it the observed fractality of the
large-scale distribution of galaxies should appear when observational
relations necessary for fractal characterization are calculated along
the past light cone in a perturbed metric of standard cosmology}.

If this conjecture proves, even partially, correct, fractals in
cosmology would no longer be necessarily seen as opposed to the
cosmological principle. Notice that this can only happen in circumstances
where fractality is characterized by an {\it observed}, smoothed out,
and averaged fractal system, as opposed to building a fractal structure
in the very spacetime geometrical structure, as initially thought
necessary to do for having fractals in cosmology (Mandelbrot 1983;
Ribeiro 1992a). Thus, the usual tools used in relativistic
cosmology, like the fluid approximation, will remain valid. As a
possible consequence of this conjecture, a detailed characterization
of the fractal structure could provide direct clues for the kind of
cosmological perturbation necessary in our cosmological models, and
this could shed more light in issues like galaxy formation.

There is now underway an attempt to check the validity of this
conjecture (Abdalla, Mohayaee, and Ribeiro 1999) by means of a specific
perturbative approach to standard cosmology (Abdalla and Mohayaee 1999).

%%%%%%%%%%%%%%%%%%%%%%%%%%%%%%%%%%%%%%%%%%%%%%%%% acknowledgements
\vspace{5mm}
Thanks go to E.\ Abdalla and R.\ Mohayaee for reading the original
manuscript and for helpful comments and suggestions. Partial support
from FUJB-UFRJ is also acknowledged.

%%%%%%%%%%%%%%%%%%%%%%%%%%%%%%%%%%%%%%%%%%%%%%%%% References
\begin{flushleft}
{\Large \bf References}
\end{flushleft}
\begin{description}
\item Abdalla, E., and Mohayaee, R.\ 1999, Phys.\ Rev.\ D, 59, 084014,
      astro-ph/9810146
\item Abdalla, E., Mohayaee, R., and Ribeiro, M.\ B.\ 1999, astro-ph/9910003
\item Amendola, L.\ 1999, preprint (to appear in the Proceedings of the IX
      Brazilian School of Cosmology and Gravitation, M.\ Novello 1999)
\item Cappi, A., Benoist, C., da Costa, L.\ N., and Maurogordato, S.\ 1998,
      A \& A, 335, 779
\item Baryshev, Y.\ V., Sylos-Labini, F., Montuori, M., Pietronero, L.,
      and Teerikorpi, P.\ 1998, Fractals, 6, 231, astro-ph/9803142
\item Charlier, C.\ V.\ L.\ 1908, Ark.\ Mat.\ Astron.\ Fys., 4, 1
\item Charlier, C.\ V.\ L.\ 1922, Ark.\ Mat.\ Astron.\ Fys., 16, 1
\item Coleman, P.\ H., and
        Pietronero, L.\ 1992, Phys.\ Rep., 213, 311
\item Coles, P.\ 1998, Nature, 391, 120
\item Davis, M.\ 1997, Critical Dialogues in Cosmology, N.\ Turok,
      Singapore: World Scientific, 1997, 13, astro-ph/9610149
\item de Vaucouleurs, G.\ 1970a, Science, 167, 1203
\item de Vaucouleurs, G.\ 1970b, Science, 168, 917
\item de Vaucouleurs, G., and Wertz, J.\ R.\ 1971, Nature, 231, 109 
\item Fournier D'Albe, E.\ E.\ 1907, Two New Worlds: I The Infra World;
      II The Supra World, London: Longmans Green
\item Longair, M.\ S.\ 1995, The Deep Universe, Saas-Fee Advanced Course 23,
      B.\ Binggeli and R.\ Buser, Berlin: Springer, 1995, 317
\item Mandelbrot, B.\ B.\ 1983, The Fractal Geometry of Nature, New
      York: Freeman
\item Mart\'{\i}nez, V.\ J.\ 1999, Science, 284, 445
\item Oldershaw, R.\ L.\ 1998, http://www.amherst.edu/\~{ }rlolders/loch.htm
\item Peebles, P.\ J.\ E. 1993, Principles of
        Physical Cosmology, Princeton University Press
\item Pietronero, L.\ 1987, Physica A, 144, 257
\item Pietronero, L., Montuori, M., and Sylos-Labini, F.\ 1997, Critical
      Dialogues in Cosmology, N.\ Turok, Singapore: World Scientific, 1997, 24 
\item Ribeiro, M.\ B.\ 1992a, Ap.\ J.\  388, 1 
\item Ribeiro, M.\ B.\ 1992b, Ap.\ J.\  395, 29 
\item Ribeiro, M.\ B.\ 1993, Ap.\ J., 415, 469 
\item Ribeiro, M.\ B.\ 1994, Deterministic Chaos in General Relativity,
      D.\ Hobbil, A.\ Burd, and A.\ Coley, New York: Plenum Press, 1994, 269
\item Ribeiro, M.\ B.\ 1995, Ap.\ J., 441, 477, astro-ph/9910145 
\item Ribeiro, M.\ B., and Miguelote, A.\ Y.\ 1998, Brazilian J.\ Phys.,
      28, 132, astro-ph/9803218
\item Sandage, A., Tammann, G.\ A., and Hardy, E.\ 1972, ApJ, 172, 253
\item Sylos-Labini, F., Montuori, M., and Pietronero, L.\ 1998, Phys.\
      Rep., 293, 61, astro-ph/9711073
\item Wertz, J.\ R.\ 1970, Newtonian Hierarchical Cosmology, PhD thesis
      (University of Texas at Austin, 1970)
\item Wertz, J.\ R.\ 1971, Ap.\ J., 164, 227
\item Wu, K.\ K.\ S., Lahav, O., and Rees, M.\ J.\ 1999, Nature, 397,
      225, astro-ph/9804062
\end{description}
\end{document}